\documentclass[12pt,english]{article}
\usepackage[T1]{fontenc}
\usepackage{a4wide}
\usepackage{babel}
\usepackage{graphics}

\makeatletter

\providecommand{\LyX}{L\kern-.1667em\lower.25em\hbox{Y}\kern-.125emX\@}
\newcommand{\noun}[1]{\textsc{#1}}

\usepackage{amssymb}

\makeatother

\begin{document}

\renewcommand\theequation{\hbox{\normalsize\arabic{section}.\arabic{equation}}}

September 1999\hfill{}DFUB-99-16

\vfill{}
{\par\centering \textbf{\LARGE Nonlinear Integral Equation and Finite Volume
Spectrum of Minimal Models Perturbed by \( \Phi _{(1,3)} \)}\LARGE \par}
\vspace{1cm}

{\par\centering G. Feverati\footnote{
E-mail: feverati@bo.infn.it
}, F. Ravanini\footnote{
E-mail: ravanini@bo.infn.it
} and G. Tak{\' a}cs\footnote{
E-mail: takacs@bo.infn.it
}\par}
\vspace{0.5cm}

{\par\centering \emph{INFN Sezione di Bologna - Dipartimento di Fisica}\\
\emph{Via Irnerio 46}\\
\emph{40126 Bologna, Italy}\par}

\begin{abstract}
We describe an extension of the nonlinear integral equation (NLIE) method to
Virasoro minimal models perturbed by the relevant operator \( \Phi _{(1,3)} \).
Along the way, we also complete our previous studies of the finite volume spectrum
of sine-Gordon theory by considering the attractive regime and more specifically,
breather states. For the minimal models, we examine the states with zero topological
charge in detail, and give numerical comparison to TBA and TCS results. We think
that the evidence presented strongly supports the validity of the NLIE description
of perturbed minimal models.\vfill{}

\end{abstract}
\clearpage

\section{Introduction}

Finite size effects play an essential role in the investigation of (both integrable
and non-integrable) \( 1+1 \) dimensional quantum field theories QFT. They
provide a possibility to determine many important physical characteristics such
as S-matrices, mass ratios, relations between parameters appearing in the ultraviolet
and infrared descriptions of the theory and a great deal of qualitative information
on the spectrum. Besides their applications in the study of low-dimensional
QFT, they appear naturally in statistical physics and in the context of lattice
field theory as well.

In this paper we will focus on integrable theories. One of the first approaches
to compute finite size effects in integrable theories was the Thermodynamic
Bethe Ansatz (TBA) \cite{YY} which was used to calculate the vacuum (Casimir)
energy \cite{Zam-tba1}. The method was later extended to include ground states
of charged sectors \cite{Fendley}. More recently, using analytic properties
of the TBA equations extended for complex values of the volume parameter, an
approach to get excited states was proposed in \cite{tateo_dorey}. Their method
to get excited states sheds light on the analytic structure of the dependence
of scaling functions on the spatial volume and up to now was the only method
developed to deal with excited states in perturbations of minimal models. Its
main drawbacks are that (1) it can be used only for systems for which a TBA
equation describing the vacuum in finite volume is known; (2) to obtain the
equation for a given excited state one has to do analytic continuation for each
case separately, and a major part of this continuation can only be carried out
numerically. Because of the requirement of the knowledge of the vacuum TBA equation
and the complications of the analytic continuation, this method is limited at
present to simple cases of integrable perturbations of Virasoro minimal models
and some other perturbed conformal field theories. Similar results were obtained
in \cite{blz, susy}.

This paper reports on a novel approach to the excited states of RCFTs in finite
volume, based on the \emph{nonlinear integral equation} (NLIE) method, which
has its origin in the so-called \emph{light-cone lattice Bethe Ansatz} approach
to regularize integrable QFTs. It was argued in \cite{ddv-87} that sine-Gordon
theory can be regularized using an inhomogeneous \( 6 \)-vertex model (or equivalently,
an inhomogeneous \( XXZ \) chain). The NLIE was originally developed in this
framework to describe the ground state scaling function (Casimir energy) in
sine-Gordon theory in \cite{ddv-92,ddv-95} and it was shown that in the ultraviolet
limit it reproduces the correct value of the central charge \( c=1 \). We remark
that similar methods were independently introduced in Condensed Matter Physics
by other authors \cite{klumper}. 

The NLIE was first extended to excited states in \cite{fioravanti} where the
spectrum of states containing only solitons (and no antisolitons/breathers)
has been described. Using an idea by Zamolodchikov \cite{plymer} they also
showed that a twisted version of the equation was able to describe ground states
of unitary Virasoro minimal models perturbed by the operator \( \Phi _{(1,3)} \).
A framework for generic excited states of even topological charge in sine-Gordon
theory was outlined by Destri and de Vega in \cite{DdV-97}. However, there
has been a contradiction between the results of the two papers, which was resolved
in \cite{our_letter1, our_npb} where we showed that it was related to the locality
and the operator content of limiting ultraviolet conformal field theory (CFT).
Besides that, we gave strong evidence for the correctness of the predicted spectrum
by comparing it to predictions coming from the truncated conformal space (TCS)
method, pioneered by Yurov and Zamolodchikov in \cite{yurov-zamolodchikov}
and extended to \( c=1 \) theories in \cite{our_letter1, our_npb}. Later we
conjectured a modification of the NLIE to describe the states of sine-Gordon/massive
Thirring theory with odd topological charge \cite{our_letter2}.

The NLIE for sine-Gordon theory was generalized to models built on general simply-laced
algebras of \( ADE \) type in \cite{mariottini} for the case of the vacuum.
More recently, in \cite{zinn-justin} P. Zinn-Justin extended the method to
the spectrum of excited states for these models and he also made a first attempt
to describe perturbations of minimal models of CFT. We will return to discussing
his results and their relations to the present paper in section 4.

The purpose of the present paper is to present a framework for describing general
excited states of Virasoro minimal models (including nonunitary ones) perturbed
by the operator \( \Phi _{(1,3)} \) (we consider only the massive case). In
order to do that first we extend our previous studies to describe breather states
in section 2. We show that in the IR limit the scaling functions resulting from
the NLIE match the S-matrices conjectured in \cite{zamzam} and then proceed
to a numerical comparison to TCS data. We also discuss some consequences coming
from the evaluation of the conformal weights of the UV limiting states.

In section 3 we give a description of ground states of the perturbed minimal
models and check them against TCS and, where available, TBA data. The comparison
with TBA is especially powerful, since the numbers coming from the TBA method
are exact up to the numerical precision of the iterative solution (which can
be as small as computing power allows). Then in section 4 we write down and
discuss the NLIE for the excited states. In our exposition we restrict ourselves
to neutral states. In section 5 we check the simplest examples of the resulting
excited state energies both numerically and qualitatively against TCS, and in
section 6 we give our conclusions.

\section{The NLIE for the sine-Gordon theory in the attractive regime}

\subsection{Notations and conventions}

Our convention for the sine-Gordon Lagrangian is
\begin{equation}
\label{sG_Lagrangian}
{\cal L}_{sG}=\displaystyle\int \left( \displaystyle\frac{1}{2}\partial _{\nu }\Phi \partial ^{\nu }\Phi +\lambda :\cos \left( \beta \Phi \right) :\right) dx\, ,
\end{equation}
where \( \beta  \) is the coupling constant and the dimensionful parameter
\( \lambda  \) essentially defines a mass scale, which can be expressed in
terms of the soliton mass \( M \) \cite{mass_scale} as follows:
\begin{equation}
\label{mass_scale_sG}
\lambda =\displaystyle\frac{2\Gamma (\Delta )}{\pi \Gamma (1-\Delta )}\left( \displaystyle\frac{\sqrt{\pi }\Gamma \left( \displaystyle\frac{1}{2-2\Delta }\right) }{2\Gamma \left( \displaystyle\frac{\Delta }{2-2\Delta }\right) }M\right) ^{2-2\Delta }\, \, \, ,\, \, \Delta =\displaystyle\frac{\beta ^{2}}{8\pi }\, \, .
\end{equation}
 For later convenience, we define a new parameter \( p \) by 
\begin{equation}
\label{p_definition}
p=\displaystyle\frac{\beta ^{2}}{8\pi -\beta ^{2}}\: .
\end{equation}
In the repulsive regime \( p>1 \) while in the attractive \( p<1 \). The free
fermion point is at \( p=1 \) and the \( k \)th breather threshold is \( p=1/k \).
For later reference we recall that the UV limit of the theory is a \( c=1 \)
CFT, whose spectrum of primary fields consists of vertex operators \( V_{n,m} \)
with the conformal weights
\begin{equation}
\label{coulombgas}
\Delta ^{\pm }=\displaystyle\frac{1}{2}\left( \displaystyle\frac{n}{R}\pm \displaystyle\frac{1}{2}mR\right) ^{2}\, \, ,
\end{equation}
where \( R \) is the compactification radius of the \( c=1 \) free boson,
related to \( p \) by
\[
\frac{1}{2R^{2}}=\frac{p}{p+1}\, \, .\]

Let us briefly recall the NLIE for sine-Gordon theory. We will use the notations
and conventions of the paper \cite{our_npb}, which the reader is invited to
consult for more details. We put the sine-Gordon model on a cylindrical spacetime,
with the infinite time direction and compact spatial extension of length (volume)
\( L \). The NLIE is a complex nonlinear integral equation for the \emph{counting
function \( Z(\vartheta ) \)}:
\begin{equation}
\label{nlie-cont}
\displaystyle \begin{array}{cc}
Z(\vartheta )=l\sinh \vartheta +g(\vartheta |\vartheta _{j})+C & -i\displaystyle\int ^{\infty }_{-\infty }dxG(\vartheta -x-i\eta )\log \left( 1+(-1)^{\delta }e^{iZ(x+i\eta )}\right) \\
 & +i\displaystyle\int ^{\infty }_{-\infty }dxG(\vartheta -x+i\eta )\log \left( 1+(-1)^{\delta }e^{-iZ(x-i\eta )}\right) 
\end{array}
\end{equation}
where \( l=ML \) is the dimensionless volume parameter and the parameter \( \delta  \)
takes the values \( 0 \) or \( 1 \). The kernel of the equation is given by
\[
G(\vartheta )=\frac{1}{2\pi }\int ^{+\infty }_{-\infty }dk\, e^{ik\vartheta }\frac{\sinh \frac{\pi (p-1)k}{2}}{2\sinh \frac{\pi pk}{2}\, \cosh \frac{\pi k}{2}}\, .\]
The function \( g(\vartheta |\vartheta _{j}) \) is the so-called \emph{source
term}, composed of the contributions from the holes, special objects (roots/holes)
and complex roots which we call \emph{sources} and denote their positions by
the general symbol \( \{\vartheta _{j}\}=\{h_{k}\, ,\, y_{k}\, ,\, c_{k}\, ,\, w_{k}\} \)
(\( h \) stands for holes, \( y \) for special objects and \( c \) (\( w \))
for close (wide) complex roots). The source term takes the general form
\[
\displaystyle g(\vartheta |\vartheta _{j})=\sum ^{N_{H}}_{k=1}\chi (\vartheta -h_{k})-2\sum ^{N_{S}}_{k=1}\chi (\vartheta -y_{k})-\sum ^{M_{C}}_{k=1}\chi (\vartheta -c_{k})-\sum ^{M_{W}}_{k=1}\chi (\vartheta -w_{k})_{II}\: ,\]
where 
\begin{equation}
\label{chi}
\chi (\vartheta )=2\pi \displaystyle\int _{0}^{\vartheta }dxG(x)
\end{equation}
and the \emph{second determination} for any function \( f(\vartheta ) \) is
defined by 
\begin{equation}
\label{2nd_determination}
f(\vartheta )_{II}=\left\{ \begin{array}{ll}
f(\vartheta )+f\left( \vartheta -i\pi \mathrm{sign}\left( \Im m\vartheta \right) \right) \: , & p>1\: ,\\
f(\vartheta )-f\left( \vartheta -ip\pi \mathrm{sign}\left( \Im m\vartheta \right) \right) \: , & p<1\: .
\end{array}\right. \: .
\end{equation}

The source positions \( \vartheta _{i} \) are determined from the \emph{Bethe
quantization conditions}
\begin{equation}
\label{quantum}
Z(\vartheta _{j})=2\pi I_{j}\, \, ,\, \, I_{j}\in {\mathbb Z}+\displaystyle\frac{1-\delta }{2}\, \, ,
\end{equation}
where \( I_{j} \) are the Bethe quantum numbers. A complex root \( \vartheta _{j} \)
is called \emph{close} if \( |\Im m\vartheta _{j}|<\min (\pi ,\pi p) \) , otherwise
it is \emph{wide}. Complex roots always come in complex conjugate pairs, except
for self-conjugate roots which satisfy \( \Im m\vartheta _{j}=\pm \pi \displaystyle\frac{p+1}{2} \).
Special roots/holes are real positions \( y_{j} \) where the Bethe quantization
condition (\ref{quantum}) is satisfied, but the counting function is decreasing,
i.e. 
\[
Z'\left( y_{j}\right) <0.\]
Normally, \( Z \) is a monotonous increasing function on the real axis. 

\( C \) is an integration constant, which is a multiple of \( \pi  \) and
can always be set to zero by an appropriate redefinition of the counting function
\( Z \) and the source term \( g \) (see \cite{our_npb}). The redefinition
affects only self-conjugate roots and may change the Bethe quantum numbers \( I_{j} \)
from integer (i.e. \( \delta =1 \)) to half-integer (i.e. \( \delta =0 \))
or vice versa.

The NLIE (\ref{nlie-cont}) only determines \( Z(\vartheta ) \) in the fundamental
analyticity strip \( |\Im m\vartheta |<\min (\pi ,\pi p) \). Outside the strip
one must use the second determination of the counting function given by
\begin{equation}
\label{nlie-2nd}
\displaystyle \begin{array}{cc}
Z(\vartheta )_{II}=l\sinh (\vartheta )_{II}+g(\vartheta |\vartheta _{j})_{II}+C_{II} & -i\displaystyle\int ^{\infty }_{-\infty }dxG(\vartheta -x-i\eta )_{II}\log \left( 1+(-1)^{\delta }e^{iZ(x+i\eta )}\right) \\
 & +i\displaystyle\int ^{\infty }_{-\infty }dxG(\vartheta -x+i\eta )_{II}\log \left( 1+(-1)^{\delta }e^{-iZ(x-i\eta )}\right) \, \, ,
\end{array}
\end{equation}
where \( C_{II} \) is yet another integration constant, which can be set to
zero, similarly to \( C \), by redefining the source terms appropriately. In
addition, the Bethe Ansatz is periodic with a period \( i\pi (p+1) \), and
so a fundamental domain for \( Z \) can be chosen as 
\[
-\frac{\pi (p+1)}{2}<\Im m\vartheta \leq \frac{\pi (p+1)}{2}\, .\]
The two determinations can be shown to suffice to cover the fundamental domain,
outside of which \( Z \) is determined by periodicity. The counting function
takes real values on the real axis and on the boundary lines of the fundamental
domain, the so-called \emph{self-conjugate lines}.

The topological charge \( Q \) of any state can be obtained from the \emph{counting
equation}
\begin{equation}
\label{conteggi}
N_{H}-2N_{S}=2S+M_{C}+2\theta (p-1)M_{W}\: ,
\end{equation}
where \( N_{H} \) is the total number of holes (ordinary/special), \( N_{S} \)
is the number of special roots/holes, \( M_{C} \) is the number of close complex
roots and \( M_{W} \) is the number of wide roots. \( S \) is called the total
spin in the lattice Bethe Ansatz and is related to the topological charge by
\( Q=2S \). It is known that \( N_{H}-2N_{S} \) gives the number of solitonic/antisolitonic
particles of the state \cite{DdV-97}. In the repulsive regime, the complex
roots describe the internal degrees of freedom (polarization states) of solitons.
In the attractive regime, however, configurations consisting entirely of wide
roots describe the breathers and it is clear from eqn. (\ref{conteggi}) that
they do not contribute to the topological charge.

The energy and the momentum of a state can be expressed as
\begin{equation}
\label{energy}
\begin{array}{rl}
\displaystyle E-E_{bulk} & =M\displaystyle\sum ^{N_{H}}_{j=1}\cosh h_{j}-2M\displaystyle\sum ^{N_{S}}_{j=1}\cosh y_{j}\\
 & -M\displaystyle\sum ^{M_{C}}_{j=1}\cosh c_{j}-M\displaystyle\sum _{j=1}^{M_{W}}(\cosh w_{j})_{II}\\
 & -M\displaystyle\int ^{\infty }_{-\infty }\displaystyle\frac{dx}{2\pi }2\Im m\left[ \sinh (x+i\eta )\log (1+(-1)^{\delta }e^{iZ(x+i\eta )})\right] \: ,
\end{array}
\end{equation}
\begin{equation}
\label{momentum}
\begin{array}{rl}
\displaystyle P & =M\displaystyle\sum ^{N_{H}}_{j=1}\sinh h_{j}-2M\displaystyle\sum ^{N_{S}}_{j=1}\sinh y_{j}\\
 & -M\displaystyle\sum ^{M_{C}}_{j=1}\sinh c_{j}-M\displaystyle\sum _{j=1}^{M_{W}}(\sinh w_{j})_{II}\\
 & -M\displaystyle\int ^{\infty }_{-\infty }\displaystyle\frac{dx}{2\pi }2\Im m\left[ \cosh (x+i\eta )\log (1+(-1)^{\delta }e^{iZ(x+i\eta )})\right] \: ,
\end{array}
\end{equation}
where the values of \( h_{j},\, c_{j},\, y_{j} \) and \( w_{j} \) are fixed
by the quantization conditions (\ref{quantum}). (Note that the sign of the
wide root contribution is different from that in \cite{our_npb} - in our previous
paper there is a misprint). The \emph{bulk energy term} takes the form \cite{mass_scale}
\begin{equation}
\label{bulk_energy}
E_{bulk}=-\displaystyle\frac{1}{4}M^{2}L\tan \displaystyle\frac{\pi p}{2}\, \, .
\end{equation}
For \( p \) and odd integer, this is divergent and is renormalized to give
a term depending logarithmically on \( L \) (for an explanation see e.g. section
6 of \cite{our_npb}). The energy levels in finite volume take the general form
\begin{equation}
\label{energy_UV}
E(L)=-\displaystyle\frac{\pi c(L)}{6L}\, \, .
\end{equation}
\( c(L) \) is called a \emph{scaling function} and has the limiting value
\[
c(0)=c-12\left( \Delta ^{+}+\Delta ^{-}\right) \, \, ,\]
where \( \Delta ^{\pm } \) are the conformal weights of the corresponding state
in the UV limiting CFT.

\subsection{Infrared limit and breather \protect\( S\protect \)-matrices}

In the infrared limit \( l\, \rightarrow \, \infty  \) the term \( l\sinh (\vartheta ) \)
develops a large imaginary part in the first determination away from the real
axis, forcing the close complex roots to fall into special configurations called
\emph{arrays} (we use the terminology of \cite{DdV-97})\emph{.} An array is
a set of complex roots in which the roots are placed at specific intervals in
the imaginary direction and have the same real part. In the attractive regime
\( l\sinh (\vartheta )_{II} \) is nonzero and so this is true for nonselfconjugate
wide pairs as well (in the repulsive case wide roots do not have such driving
force), while self-conjugate roots have a fixed imaginary part anyway. The deviation
of the complex roots from their positions in the array decays exponentially
with \( l \) (see the paper \cite{our_npb} for an example with two holes and
one complex pair in the repulsive regime). For a more detailed discussion, see
\cite{DdV-97}.

For the rest of this subsection, whenever it is not explicitly stated, we restrict
ourselves to the attractive regime \( p<1 \). The possible arrays fall into
two classes:

\begin{enumerate}
\item \emph{Arrays of the first kind} are the ones containing close roots, which describe
the polarization states of solitons. \\
There are two degenerate cases: \emph{odd degenerate} arrays, which have a self-conjugate
root at
\[
\vartheta _{0}=\vartheta +i\frac{\pi (p+1)}{2}\]
and accompanying complex pairs at
\[
\vartheta _{k}=\vartheta \pm i\frac{\pi (1-(2k+1)p)}{2}\, \, ,\, \, k=0,\ldots ,\left[ \frac{1}{2p}\right] \]
 and \emph{even degenerate} ones, which only contain complex pairs, at the positions
\[
\vartheta _{k}=\vartheta \pm i\frac{\pi (1-2kp)}{2}\, \, ,\, \, k=0,\ldots ,\left[ \frac{1}{2p}\right] \]
These arrays always contain exactly one close pair. The odd degenerate arrays
in the repulsive regime reduce to single self-conjugate roots and the even degenerate
ones to a single close complex pair. \\
The description of the nondegenerate cases is a bit more complicated and can
be found in \cite{DdV-97}. They always contain two close complex pairs and
they are the attractive regime analogous of wide pairs in the repulsive regime,
but we will not need them here.
\item \emph{Arrays of the second kind} describe breather degrees of freedom. The odd
ones contain a self-conjugate root 
\[
\vartheta _{0}=\vartheta +i\pi (p+1)/2\]
 and wide pairs as follows:
\[
\Im m\vartheta _{k}=\vartheta \pm i\frac{\pi (1-(2k+1)p)}{2}\, \, ,\, \, k=0,\ldots ,s\, \, ,\]
where 
\[
0\leq s\leq \left[ \frac{1}{2p}\right] -1\, \, ,\]
while the even ones only contain wide pairs 
\[
\Im m\vartheta _{k}=\vartheta \pm i\frac{\pi (1-2kp)}{2}\, \, ,\, \, k=0,\ldots ,s\, \, ,\]
and \( s \) runs in the same range. They correspond to the \( (2s+1) \)-th
breather \( B_{2s+1} \) and the \( (2s+2) \)-th breather \( B_{2s+2} \),
respectively.
\end{enumerate}
As one can see, arrays of the second kind become degenerate ones of the first
kind, if we analytically continue increasing \( p \). The reason is that breathers
are of course soliton-antisoliton bound states, while degenerate arrays of the
first kind describe scattering states of a soliton and antisoliton, as we will
see shortly.

In the infrared limit one can drop all terms containing the integral of 
\[
\log (1+(-1)^{\delta }e^{\pm iZ(x\pm i\eta )})\]
because they exponentially decay with \( l \). One can therefore compute the
energy and momentum contribution of a array of the second kind corresponding
to the breather \( B_{s} \). The energy-momentum contribution turns out to
be
\begin{equation}
\label{breather-energy}
2M\sin \displaystyle\frac{\pi sp}{2}\left( \cosh \vartheta ,\sinh \vartheta \right) \, \, ,
\end{equation}
where \( \vartheta  \) is the common real part of the roots composing the array.
This is just the contribution of a breather \( B_{s} \) moving with rapidity
\( \vartheta  \). Arrays of the first kind do not contribute to the energy-momentum
in the infrared limit, which lends support to their interpretation as polarization
states of solitons.

So far we have just reviewed some fundamental facts already known in the Bethe
Ansatz literature (see \cite{DdV-97} and references therein). Now we proceed
to show that with the above interpretation the NLIE correctly reproduces the
two-body scattering matrices of sine-Gordon theory including breathers, which
has not been done before. For the repulsive regime the \( S \)-matrices were
calculated in detail in our previous paper \cite{our_npb}.

Let us start with breather-soliton matrices. The Bethe quantization conditions
for a state containing a soliton (i.e. a hole) with rapidity \( \vartheta _{1} \)
and a breather \( B_{s} \) with rapidity \( \vartheta _{2} \) take the following
form in the infrared limit. For the hole we get
\[
Z(\vartheta _{1})=M\sinh \vartheta _{1}-\sum ^{s}_{k=0}\chi (\vartheta _{1}-\vartheta _{2}-i\rho _{k})_{II}\, \, =2\pi I_{1},\]
where we denoted the prescribed imaginary parts of the roots in the array \( B_{s} \)
by \( \rho _{k} \). Here we used \( \chi (0)=0 \) to eliminate the source
term for the hole. Now we can compute the second determination of \( \chi  \)
to be 
\[
\chi (\vartheta )_{II}=\left\{ \begin{array}{cc}
gd(\vartheta +i\pi /2)+gd(\vartheta -i\pi /2+i\pi (p+1))\, \, , & \Im m\vartheta <-\pi p\\
gd(\vartheta -i\pi /2)+gd(\vartheta +i\pi /2-i\pi (p+1))\, \, , & \Im m\vartheta >\pi p
\end{array}\right. \]
where 
\[
{\rm gd}(\vartheta )=i\log \frac{\sinh (i\pi /4+\vartheta /2)}{\sinh (i\pi /4-\vartheta /2)}\, \, ,\]
which is essentially the Gudermannian \( \arctan (\sinh (\vartheta )) \) with
a suitable choice of branches \cite{DdV-97}. Now it is a matter of elementary
algebra to arrive at
\[
Z(\vartheta _{1})=M\sinh \vartheta _{1}-i\log S_{SB_{s}}(\vartheta _{1}-\vartheta _{2})=2\pi I_{1}\, \, ,\]
where \( S_{SB_{s}}(\vartheta _{1}-\vartheta _{2}) \) is the soliton-breather
\( S \)-matrix conjectured in \cite{zamzam}.

One can start with the breather quantization conditions, too. Writing 
\begin{equation}
\label{quant_br}
Z\left( \vartheta _{2}+i\rho _{k}\right) _{II}=M\sinh \left( \vartheta _{2}+i\rho _{k}\right) _{II}+\chi (\vartheta _{2}-\vartheta _{1}+i\rho _{k})_{II}+\ldots =2\pi I^{(k)}_{2}\, \, ,k=0,\ldots ,s
\end{equation}
(the dots are terms due to wide root sources themselves, which cancel out up
to multiples of \( 2\pi  \) in the next step). Summing up these equations one
arrives at
\[
2M\sin \frac{sp\pi }{2}\sinh (\vartheta _{2})-i\log S_{SB_{s}}(\vartheta _{2}-\vartheta _{1})=2\pi I_{2}\, \, ,\]
where \( I_{2} \) is essentially minus the sum of the quantum numbers of the
wide roots composing the array (shifted by some integer coming from summing
up the terms omitted in eqn. (\ref{quant_br})).

Using a similar line of argument we also reproduced the breather-breather \( S \)-matrices
by writing down the Bethe quantization conditions for a state with two degenerate
strings \( B_{s} \) and \( B_{r} \) of the second kind. One has to be careful
that when \( Z(\vartheta ) \) contains wide root sources which are expressed
in terms of \( \chi (\vartheta )_{II} \), the second determinations of these
terms will appear in \( Z(\vartheta )_{II} \), i.e. terms that can be written
roughly like \( \left( \chi (\vartheta )_{II}\right) _{II} \) .

We remark that breather \( S \)-matrices of the sine-Gordon model were also
reproduced in \cite{doikou} from the \( XXZ \) spin chain. However, their
calculation is made on a homogeneous lattice \( XXZ \) chain, while we perform
the derivation in the continuum limit of an inhomogeneous chain, which is described
by the NLIE (\ref{nlie-cont}). It is very important that we can identify the
spectral parameter \( \vartheta  \) with the rapidity of a relativistic breather
particle with the correct mass, using eqn. (\ref{breather-energy}), while in
the case of the homogeneous lattice \( XXZ \) chain the dispersion relation
is not the relativistic one. The relativistic dispersion relation (\ref{breather-energy})
can only be obtained by taking a special continuum limit, in which we send the
inhomogeneity parameter to \( \infty  \) and at the same time take the lattice
spacing to \( 0 \) (for details see \cite{DdV-97}).

Scattering state of a soliton and an antisoliton can be described by taking
two holes and a degenerate array of the first kind. There are two possibilities
now, corresponding to scattering in the parity-odd and parity-even channels.
Following the procedure outlined in \cite{our_npb}, we were once again able
to reproduce the corresponding scattering amplitudes. The results presented
here together with those of \cite{our_npb} exhaust all two-particle scattering
amplitudes of sine-Gordon theory, both in the repulsive and attractive regime.

\subsection{Some examples of breather states}

The vacuum scaling function and the multi-soliton states of sine-Gordon theory
(both in the attractive and the repulsive regime) have already been examined
in \cite{our_letter1}, where we found agreement with TCS predictions. So we
proceed to take a look at the simplest neutral excited state, which is the one
containing a first breather \( B_{1} \) at rest. The source term to be written
into the NLIE turns out to be

\begin{equation}
\label{selfconjugate_source}
g(\vartheta )=-i\log \displaystyle\frac{\cos \displaystyle\frac{\pi p}{2}-i\sinh \vartheta }{\cos \displaystyle\frac{\pi p}{2}+i\sinh \vartheta }\, \, ,
\end{equation}
and the self-conjugate root is located exactly at \( \vartheta _{0}=i\pi (p+1)/2 \),
since the breather has zero momentum. There is no need to look at the Bethe
quantization condition as the root does not move due to the left-right symmetry
of the problem. This state is quantized with integer Bethe quantum numbers,
i.e. \( \delta =1 \) (cf. \cite{our_npb, our_letter2} for an explanation of
the connection between the choice of \( \delta  \) and locality of the corresponding
quantum field theory).

Calculating the UV conformal dimensions along the lines presented in \cite{our_npb}
we obtain the conformal dimensions
\begin{equation}
\label{1stbreather_UV}
\Delta ^{\pm }=\displaystyle\frac{p}{p+1}\, \, ,
\end{equation}
so the ultraviolet limit of this state is a linear combination of the vertex
operators \( V_{\pm 1,0} \) of the \( c=1 \) UV CFT (see eqn. (\ref{coulombgas})).
This is in perfect agreement with the TCS calculations performed by us (for
an outline of the method see \cite{our_npb}). To normalize the TCS, we used
the formula (\ref{mass_scale_sG}).

Table \ref{1st_breather} presents the energy values obtained by iterating the
NLIE in comparison to results coming from TCS, at the values \( p=2/7 \) and
\( p=2/9 \).

\begin{table}
{\centering \begin{tabular}{|c|c|c|c|c|}
\hline 
\( l \)&
TCS&
NLIE&
TCS&
NLIE\\
\hline 
1.0&
2.38459&
n/a&
1.84996&
n/a\\
\hline 
1.5&
1.68168&
n/a&
1.30438&
n/a\\
\hline 
2.0&
1.35391&
n/a&
1.05038&
n/a\\
\hline 
2.5&
1.17420&
n/a&
0.91152&
n/a\\
\hline 
3.0&
1.06692&
1.0655810539&
0.82903&
0.8285879853\\
\hline 
3.5&
0.99985&
0.9980153379&
0.77773&
0.7771857432\\
\hline 
4.0&
0.95664&
0.9542867454&
0.74499&
0.7443106400\\
\hline 
4.5&
0.92845&
0.9254766107&
0.72381&
0.7229895177\\
\hline 
5.0&
0.91000&
0.9063029141&
0.71006&
0.7090838262\\
\hline 
\end{tabular}\par}

\caption{\small \label{1st_breather}The first breather state at \protect\( p=\displaystyle\frac{2}{7}\protect \)
and \protect\( p=\displaystyle\frac{2}{9}\protect \). Energies and distances are measured
in units of the soliton mass \protect\( M\protect \), and we have subtracted
the predicted bulk energy term from the TCS data.}
\end{table}
The table shows that iteration of the NLIE fails for values of \( l \) less
than \( 3 \) (the actual limiting value is around \( 2.5 \)). What is the
reason?

\begin{figure}
{\centering \begin{tabular}{ccc}
\resizebox*{0.4\columnwidth}{!}{\includegraphics{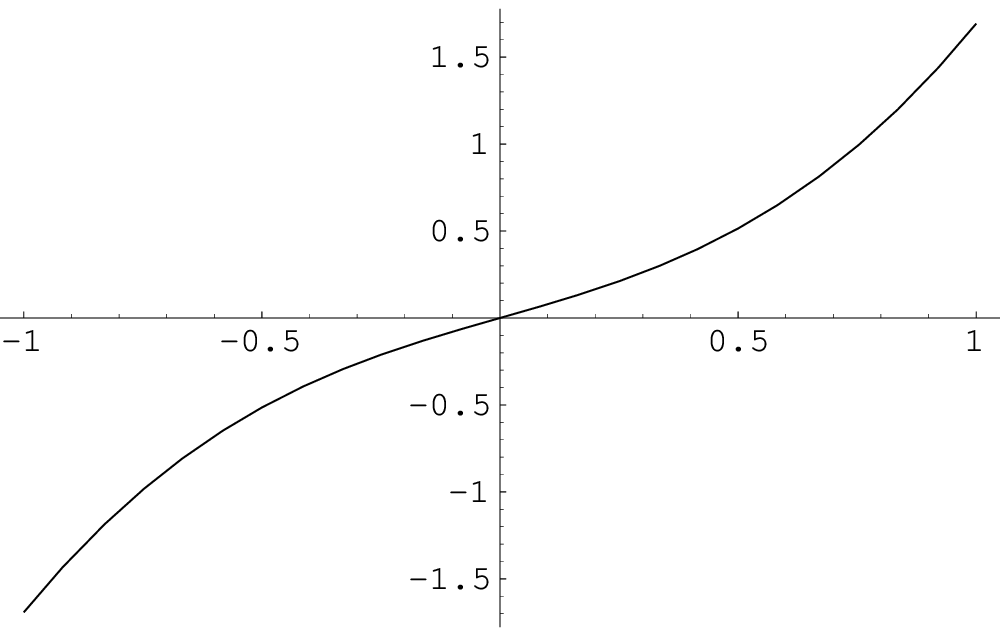}} &
&
\resizebox*{0.4\columnwidth}{!}{\includegraphics{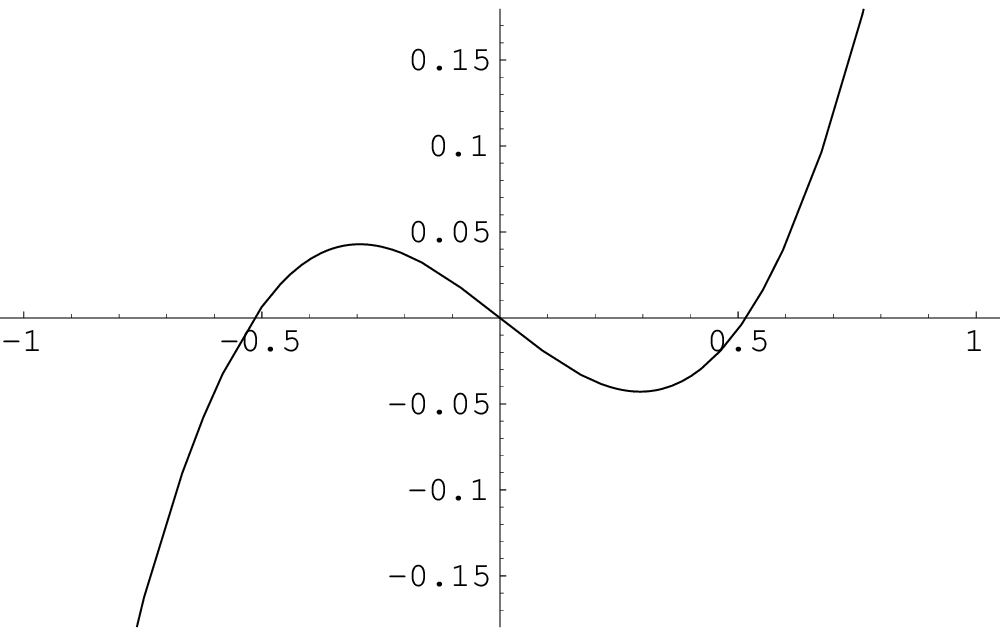}} \\
\end{tabular}\par}

\caption{\small \label{specials} The behavior of the function \protect\( Z(\vartheta )\protect \)
at \protect\( l=3\protect \) and \protect\( l=2\protect \), respectively.}
\end{figure}

We plot the counting function \( Z \) on the real line for \( l=3 \) and \( l=2 \)
in figure \ref{specials}. In a first approximation we can safely neglect the
integral term for these values of the volume to see the qualitative features
that we are interested in. 
What we see is that the behaviour of the function changes: its derivative changes
sign at the origin. As a result, two new holes appear where the new real zeros
of the function are. But the topological charge remains zero, due to the fact
that now we have a special root at the origin and so \( N_{S}=1 \) and \( N_{H}=2 \).
The two new holes do not give us any new dynamical degrees of freedom: their
quantum numbers are fixed to be \( 0 \) and so their positions are uniquely
determined.

Of course when we calculated the UV dimension, the appearance of the new sources
had to be taken into account to obtain the result (\ref{1stbreather_UV}). It
turns out that for \( \displaystyle\frac{1}{3}<p<1 \) the two holes are left/right movers,
while for \( p<\displaystyle\frac{1}{3} \) they remain central, using the terminology of
\cite{our_npb}.

How does this phenomenon affect the iteration scheme for the NLIE? The two new
zeros of \( Z(\vartheta ) \), which is a complex analytic function apart from
logarithmic branch cuts, actually correspond to singularities of the logarithmic
term in the NLIE (\ref{nlie-cont}). They come along the imaginary axis in the
\( \vartheta  \) plane as we decrease \( l \), and at a certain point they
cross our integration contour which runs parallel to the real axis at distance
\( \eta  \). As they make the logarithmic term in our NLIE (\ref{nlie-cont})
singular, they blow up our iteration scheme. After reaching the origin of the
\( \vartheta  \) plane (at exactly the radius where the derivative of \( Z \)
becomes \( 0 \)), they continue to move along the real axis, which corresponds
to crossing a square root branch cut.

We do not go into details here as this problem is currently under investigation\footnote{
Work in progress in collaboration with P. E. Dorey and C. Dunning, Durham.
}. We just remark that these issues prove to be highly nontrivial and for the
time being, unfortunately, they prevent us from having a reliable numeric scheme
for the NLIE below the critical volume. The problem will be examined in a forthcoming
publication.

One can estimate the volume where the slope of the counting function changes
sign by neglecting the integral of the logarithmic term. The result is
\begin{equation}
\label{critical_estimate}
l_{critical}=\displaystyle\frac{2}{\cos \displaystyle\frac{p\pi }{2}}
\end{equation}
which gives a value of around \( 2.22 \) for \( p=2/7 \) and \( 2.13 \) for
\( p=2/9 \). The actual limiting value is a bit higher, partly due to the finite
value of \( \eta  \) used in the iteration program and partly because the iteration
already destabilizes when the singularities come close enough to the contours.
It must also be noted that the integral term cannot eventually be neglected
when the singularities are close to the contour, which is an additional reason
why (\ref{critical_estimate}) is just a crude estimate.

We make a short digression to examine the UV limit of the second breather \( B_{2} \).
The second breather at rest is described by a wide pair at positions 
\[
\vartheta =\pm i\frac{\pi }{2}\, \, .\]
Calculating the UV conformal dimension we get 
\[
\Delta ^{+}=\Delta ^{-}=\frac{p}{p+1}\, \, ,\]
which turns out to be the same as that of the first one (\ref{1stbreather_UV}),
i.e. this state must originate from the other linearly independent combination
of the vertex operators \( V_{\pm 1,0} \) in the ultraviolet. This is again
in perfect agreement with TCS and confirms a result by Pallua and Prester \cite{pallua}
who used \( XXZ \) chain in transverse magnetic field to regularize sine-Gordon
theory. They calculated scaling functions numerically on a finite lattice for
several concrete values of \( p \), and arrived at this conclusion by looking
at the numerical data. However, our method to compute UV dimensions gives us
an \emph{exact analytic formula} and therefore much stronger evidence. This
result is interesting because it invalidates a conjecture made previously by
Klassen and Melzer \cite{heretic} who identified the second breather as a linear
combination of \( V_{\pm 2,0} \). 

To close this section, we present the lowest lying example of a two-breather
state, containing two \( B_{1} \) particles with zero total momentum. It turns
out that this is a state for which the numerical iteration of the NLIE is not
plagued with the problem found above for the first breather. Locality constrains
the state to be quantized with half-integers and for lowest energy the quantum
numbers of the self-conjugate roots must take the values
\[
I_{1}=\frac{1}{2}\, \, ,\, \, I_{2}=-\frac{1}{2}\]
We remark that in contrast to the case of holes, the self-conjugate root with
\( I>0 \) moves to the left, while the one with \( I<0 \) moves to the right.
This is due to the fact that the second determination of \( Z \) is in general
a monotonically decreasing function on the self-conjugate line. In order to
determine the position of the two self-conjugate roots we need the second determination
of \( Z \). The second determination of the self-conjugate root source turns
out to be

\[
\left( \chi (\vartheta )_{II}\right) _{II}=i\log \frac{i\sin \pi p-\sinh \vartheta }{i\sin \pi p+\sinh \vartheta }\, \, .\]
Up to some signs, this is just the phase shift which arises when two breathers
scatter on each other, which is exactly why the IR analysis gives the correct
scattering amplitude. Using this formula, we obtained the numerical data presented
in table \ref{two_breathers}.
\begin{table}
{\centering \begin{tabular}{|c|c|c|}
\hline 
\( l \)&
TCS&
NLIE\\
\hline 
1.0&
12.1601&
12.159257\\
\hline 
1.5&
8.20139&
8.2006130\\
\hline 
2.0&
6.24771&
6.2465898\\
\hline 
2.5&
5.09489&
5.0937037\\
\hline 
3.0&
4.34132&
4.3397021\\
\hline 
3.5&
3.81513&
3.8129798\\
\hline 
4.0&
3.43020&
3.4275967\\
\hline 
4.5&
3.13912&
3.1357441\\
\hline 
5.0&
2.91308&
2.9089439\\
\hline 
\end{tabular}\par}

\caption{\small \label{two_breathers} The two-breather state at \protect\( p=\frac{2}{7}\protect \).}
\end{table}
The UV limit for this state can be calculated from NLIE to be a symmetric first
level descendent of the vacuum with weights
\[
\Delta ^{+}=\Delta ^{-}=1\, \, ,\]
which agrees with TCS. (Note that this descendent exists due to the fact that
there is a \( \hat{U}(1)_{L}\times \hat{U}(1)_{R} \) Kac-Moody symmetry at
\( c=1 \): this state exactly corresponds to the combination of the left and
right moving currents \( J\bar{J} \).)

We remark that the precision of \( c=1 \) TCS is at maximum \( 5-6 \) digits
for low lying states and small values of \( l \), and the truncation error
increases with the volume and as we go higher and higher in the spectrum. We
also mention that to achieve this precision it is necessary to go up to around
\( 4000-5000 \) states, which is a highly nontrivial task accomplished only
by writing all the calculation in compiled \( C \) programs (including the
evaluation of CFT matrix elements) and it was really stretching the computing
power available to us to its limits. In contrast, for states without moving
roots the NLIE can be readily iterated up to \( 12 \) digits precision by a
simple personal computer and even for states with moving roots like the two-breather
one it is not very hard to achieve a precision of \( 6-7 \) digits (although
the number of necessary iterations grows when one decreases the volume parameter
\( l \) and the prescribed precision).

\section{Zamolodchikov's \protect\( \alpha \protect \)-twist and ground states for
minimal models perturbed by \protect\( \Phi _{(1,3)}\protect \)}

Some time ago, Zamolodchikov has put forward the idea of modifying sine-Gordon
theory by a twist \( \alpha  \) \cite{plymer}. The NLIE describing this situation
has the following form:
\begin{equation}
\label{twisted-nlie}
\begin{array}{ll}
Z(\vartheta )=l\sinh \vartheta +\alpha -i & \displaystyle\int ^{\infty }_{-\infty }dxG(\vartheta -x-i\eta )\log \left( 1+e^{iZ(x+i\eta )}\right) \\
 & +i\displaystyle\int ^{\infty }_{-\infty }dxG(\vartheta -x+i\eta )\log \left( 1+e^{-iZ(x-i\eta )}\right) 
\end{array}
\end{equation}
For definiteness we choose half-integer quantization rule with \( \delta =0 \),
since it is obvious that shifting \( \alpha  \) by \( \pi  \) and then redefining
\( Z \) one can change the Bethe quantum numbers from half-integers to integers.
The energy level determined by this equation has the leading UV behavior 

\[
E(L)=-\frac{\pi \tilde{c}}{6L}+\ldots \]
with

\begin{equation}
\label{alpha_vacuum}
\tilde{c}=1-\displaystyle\frac{6p}{p+1}\left( \displaystyle\frac{\alpha }{\pi }\right) ^{2}\, \, .
\end{equation}
Furthermore, it is well-known that the perturbation of the Virasoro minimal
model \( Vir(r,s) \) by its relevant primary operator \( \Phi _{(1,3)} \)
is integrable and is described by an RSOS restriction of sine-Gordon theory
\cite{reshetikhin-smirnov} with 
\begin{equation}
\label{p_min_r_s}
p=\displaystyle\frac{r}{s-r}\, \, .
\end{equation}
We will use for this model the shorthand notation \( Vir(r,s)+\Phi _{(1,3)} \).

Now, putting \( \alpha =\pi /r \)~~ we get

\[
\tilde{c}=1-\frac{6}{rs}\, \, ,\]
which is exactly the \emph{effective central charge} of the minimal model \( Vir(r,s) \).
Therefore one can expect that the twisted equation describes the ground state
of the model \( Vir(r,s)+\Phi _{(1,3)} \). In fact, Fioravanti et al. \cite{fioravanti}
calculated these scaling functions for the unitary case \( s=r+1 \) and showed
that they match perfectly with the TBA predictions already available . Moreover,
choosing the following values for the twist
\begin{equation}
\label{twists}
\alpha =\pm \displaystyle\frac{k\pi }{r}\, \, ,\, \, k=1\ldots r-1\, \, 
\end{equation}
they obtained the conformal weights of the operators \( \Phi _{(k,k)}\, \, ,\, \, k=1\ldots r-1 \)
in the UV limit (the sign choice is just a matter of convention). In our notation,
\( \Phi _{(k,l)} \) denotes the primary field with conformal weights 
\begin{equation}
\label{Kac_formula}
\Delta ^{+}=\Delta ^{-}=\displaystyle\frac{(ks-lr)^{2}-(s-r)^{2}}{4sr}\, \, .
\end{equation}
The models \( Vir(r,s)+\Phi _{(1,3)} \) have exactly \( r-1 \) ground states.
In fact, one can see from the fusion rules that the matrix of the operator \( \Phi _{(1,3)} \)
is block diagonal with exactly \( r-1 \) blocks in the Hilbert space made up
of states with the same left and right primary weights. In each of these blocks,
there is exactly one ground state and for the unitary series \( s=r+1 \), it
was conjectured in \cite{gr_states} that their UV limits are the states corresponding
to \( \Phi _{(k,k)} \). One can check that in the general nonunitary case the
twists (\ref{twists}) correspond in the UV limit to the lowest dimension operators
among each of the \( r-1 \) different blocks of primaries (see explicit examples
later). These ground states are degenerate in infinite volume, but for finite
\( l \) they are split with their gaps decaying exponentially as \( l\, \rightarrow \, \infty  \).
In the unitary case, they were first analyzed in the context of the NLIE in
\cite{fioravanti} where it was shown that the NLIE predictions perfectly match
with the TBA results already available for the unitary series.

However, ground states for nonunitary models have not been treated so far and
therefore now we proceed to give examples of ground state calculations for \emph{nonunitary}
models. The models we select are the ones that will be used for comparison in
the case of excited states as well. The first is for the scaling Lee-Yang model
\( Vir(2,5)+\Phi _{(1,3)} \), for which we have also given data from TCS \cite{yurov-zamolodchikov}
and TBA \cite{Zam-tba1} for comparison (table \ref{25_vacuum}). The notation
used is 
\[
l_{B}=M_{B}L\, \, ,\]
where 
\[
M_{B}=2M\sin \frac{\pi p}{2}=\sqrt{3}M\]
 is the mass of the fundamental particle of the Lee-Yang model (this is more
natural here than using the mass \( M \) of the soliton of the unrestricted
sine-Gordon model as a scale, since the soliton disappears entirely from the
spectrum after RSOS restriction). There is only one independent value of the
twist, which we choose to be 
\[
\alpha =\frac{\pi }{2}\, \, .\]
Here and in all other subsequent calculations the TCS data were normalized using
the analogue of the coupling-mass gap relation (\ref{mass_scale_sG}) from \cite{mass_scale}.
\begin{table}
{\centering \begin{tabular}{|c|c|c|c|}
\hline 
\( l_{B} \)&
TCS&
NLIE&
TBA\\
\hline 
0.1&
-2.0835015786&
-2.0835015787&
-2.0835015786\\
\hline 
0.5&
-0.3803475256&
-0.3803475281&
-0.3803475281\\
\hline 
1.0&
-0.1532068463&
-0.1532068801&
-0.1532068801\\
\hline 
1.5&
-0.0763483319&
-0.0763484842&
-0.0763484842\\
\hline 
2.0&
-0.0406269362&
-0.0406273676&
-0.0406273676\\
\hline 
2.5&
-0.0222292932&
-0.0222302407&
-0.0222302407\\
\hline 
3.0&
-0.0123492438&
-0.0123510173&
-0.0123510173\\
\hline 
3.5&
-0.0069309029&
-0.0069338817&
-0.0069338817\\
\hline 
4.0&
-0.0039198117&
-0.0039244430&
-0.0039244430\\
\hline 
5.0&
-0.0012721417&
-0.0012816882&
-0.0012816882\\
\hline 
\end{tabular}\par}

\caption{\small \label{25_vacuum} The vacuum of the Virasoro minimal model \protect\( Vir(2,5)\protect \)
perturbed by \protect\( \Phi _{(1,3)}\protect \). The energy and the volume
are normalized to the mass of the lowest excitation, which is the first breather
of the unrestricted sine-Gordon model. The TCS data shown have the predicted
bulk energy term subtracted.}
\end{table}
There is only one ground state in this model, which corresponds to the primary
field with conformal weights 
\[
\Delta ^{+}=\Delta ^{-}=-\frac{1}{5}\, \, ,\]
which is in agreement with TBA and TCS predictions. We have also found a perfect
agreement for the models \( Vir(2,7)+\Phi _{(1,3)} \) and \( Vir(2,9)+\Phi _{(1,3)} \),
but we do not present those data here. We remark that the TCS for the minimal
models converges much better than the one for \( c=1 \) theories: all TCS data
in table \ref{25_vacuum} and subsequent ones were produced by taking a few
hundred states and in some fortunate cases (e.g. the ground state of the scaling
Lee-Yang model for small values of \( l \)) we were able to produce data with
up to \( 9-10 \) digits of accuracy! The better convergence meant that all
the computation could be done with the computer algebra program \emph{Mathematica},
greatly simplifying the programming work.

Models of the class \( Vir(2,2n+1)+\Phi _{(1,3)} \) have only one ground state.
For models with two ground states, we can take a look at \( Vir(3,5) \) and
\( Vir(3,7) \). In the first case the ultraviolet spectrum is defined by the
following Kac table, where the weight (\ref{Kac_formula}) of the field \( \Phi _{(k,l)} \)
is found in the \( k \)-th row and \( l \)-th column.

\vspace{0.3cm}
{\centering \begin{tabular}{|c|c|c|c|}
\hline 
\( 0 \)&
\( -\frac{1}{20} \)&
\( \frac{1}{5} \)&
\( \frac{3}{4} \)\\
\hline 
\( \frac{3}{4} \)&
\( \frac{1}{5} \)&
\( -\frac{1}{20} \)&
\( 0 \)\\
\hline 
\end{tabular}\par}
\vspace{0.3cm}

The two blocks of the perturbing operator \( \Phi _{(1,3)} \) are defined by
the fields \( \{\Phi _{(1,2)}\, ,\, \Phi _{(1,4)}\} \) and \( \{\Phi _{(1,1)}\, ,\, \Phi _{(1,3)}\} \),
respectively. The ground states correspond in the UV to the operators \( \Phi _{(1,2)} \)
and \( \Phi _{(1,1)} \) , as can be checked directly using formulae (\ref{alpha_vacuum}),
(\ref{p_min_r_s}) and (\ref{twists}).

We also have TBA data to compare with, using the TBA equation written by Christe
and Martins \cite{christe-martins}. The lower-lying ground state is obtained
directly from their TBA, while for the other we used Fendley's idea of twisting
the TBA equation \cite{Fendley}. The numerical results are presented in tables
\ref{vacuum_35_1} and \ref{vacuum_35_2}.

In the case \( Vir(3,7) \) (table \ref{vacua_37}) we can only have a comparison
with TCS results, but it still looks pretty convincing. The spectrum of the
UV theory is 

\vspace{0.3cm}
{\centering \begin{tabular}{|c|c|c|c|c|c|}
\hline 
\( 0 \)&
\( -\frac{5}{28} \)&
\( -\frac{1}{7} \)&
\( \frac{3}{28} \)&
\( \frac{4}{7} \)&
\( \frac{5}{4} \)\\
\hline 
\( \frac{5}{4} \)&
\( \frac{4}{7} \)&
\( \frac{3}{28} \)&
\( -\frac{1}{7} \)&
\( -\frac{5}{28} \)&
\( 0 \)\\
\hline 
\end{tabular}\par}
\vspace{0.3cm}

Here the two blocks are \( \{\Phi _{(1,2)}\, ,\, \Phi _{(1,4)}\, ,\, \Phi _{(1,6)}\} \)
and \( \{\Phi _{(1,1)}\, ,\, \Phi _{(1,3)}\, ,\, \Phi _{(1,5)}\} \), respectively.
The ground states correspond in the UV to the operators \( \Phi _{(1,2)} \)
and \( \Phi _{(1,3)} \) .
\begin{table}
{\centering \begin{tabular}{|c|c|c|c|}
\hline 
\( l \)&
TCS&
NLIE&
TBA\\
\hline 
0.1&
-3.074916&
-3.0749130189&
-3.0749130190\\
\hline 
0.3&
-0.944161&
-0.9441276204&
-0.9441276204\\
\hline 
0.5&
-0.509764&
-0.5096602194&
-0.5096602194\\
\hline 
0.8&
-0.265436&
-0.2651431026&
-0.2651431026\\
\hline 
1.0&
-0.186038&
-0.1855606546&
-0.1855606546\\
\hline 
1.5&
-0.087300&
-0.0861426792&
-0.0861426792\\
\hline 
2.0&
-0.045910&
-0.0437473815&
-0.0437473815\\
\hline 
2.5&
-0.026746&
-0.0232421927&
-0.0232421927\\
\hline 
3.0&
-0.017868&
-0.0126823057&
-0.0126823057\\
\hline 
4.0&
-0.013546&
-0.0039607326&
-0.0039607326\\
\hline 
\end{tabular}\par}

\caption{\small \label{vacuum_35_1} One of the two ground states of the Virasoro minimal model
\protect\( Vir(3,5)\protect \) perturbed by \protect\( \Phi _{(1,3)}\protect \),
corresponding to \protect\( \alpha =\frac{\pi }{3}\protect \). The energy and
the volume are normalized to the mass of the lowest excitation, which is the
soliton of the unrestricted sine-Gordon model. The TCS data shown have the predicted
bulk energy term subtracted.}
\end{table}
\begin{table}
{\centering \begin{tabular}{|c|c|c|c|}
\hline 
\( l \)&
TCS&
NLIE&
TBA\\
\hline 
0.1&
3.117844&
3.1178476855&
3.1178476853\\
\hline 
0.3&
0.985360&
0.9853990810&
0.9853990810\\
\hline 
0.5&
0.540427&
0.5405470784&
0.5405470784\\
\hline 
0.8&
0.282725&
0.2830552991&
0.2830552991\\
\hline 
1.0&
0.197143&
0.1976769278&
0.1976769278\\
\hline 
1.5&
0.089277&
0.0905539780&
0.0905539780\\
\hline 
2.0&
0.042960&
0.0453290013&
0.0453290013\\
\hline 
2.5&
0.019978&
0.0238075022&
0.0238075022\\
\hline 
3.0&
0.007209&
0.0128843786&
0.0128843786\\
\hline 
4.0&
-0.006592&
0.0039866371&
0.0039866371\\
\hline 
\end{tabular}\par}

\caption{\small \label{vacuum_35_2} The other ground state of the Virasoro minimal model \protect\( Vir(3,5)\protect \)
perturbed by \protect\( \Phi _{(1,3)}\protect \), corresponding to \protect\( \alpha =\frac{2\pi }{3}\protect \).
The energy and the volume are normalized to the mass of the lowest excitation,
which is the soliton of the unrestricted sine-Gordon model. The TCS data shown
have the predicted bulk energy term subtracted.}
\end{table}

\begin{table}
{\centering \begin{tabular}{|c|c|c|c|c|}
\hline 
&
\multicolumn{2}{|c|}{ground state 1}&
\multicolumn{2}{|c|}{ground state 2}\\
\hline 
\( l \)&
TCS&
NLIE&
TCS&
NLIE\\
\hline 
0.1&
-3.7039095315&
-3.7039095318&
0.7770591073&
0.7770591068\\
\hline 
0.5&
-0.6377729020&
-0.6377729703&
0.2075541041&
0.2075539646\\
\hline 
1.0&
-0.2314222239&
-0.2314229899&
0.1136501615&
0.1136487561\\
\hline 
1.5&
-0.1038698195&
-0.1038727861&
0.0649219172&
0.0649170326\\
\hline 
2.0&
-0.0505812388&
-0.0505886086&
0.0368105709&
0.0367993127\\
\hline 
2.5&
-0.0258481751&
-0.0258625389&
0.0208408686&
0.0208198502\\
\hline 
3.0&
-0.0136578583&
-0.0136819143&
0.0118382379&
0.0118037311\\
\hline 
3.5&
-0.0073877811&
-0.0742408962&
0.0067682934&
0.0067164173\\
\hline 
4.0&
-0.0040552779&
-0.0410602666&
0.0039103368&
0.0038372749\\
\hline 
5.0&
-0.0012235029&
-0.0013070235&
0.0013931394&
0.0012676541\\
\hline 
\end{tabular}\par}

\caption{\small \label{vacua_37} The two ground states of the Virasoro minimal model \protect\( Vir(3,7)\protect \)
perturbed by \protect\( \Phi _{(1,3)}\protect \). The energy and the volume
are normalized to the mass of the kink, which is the soliton of the unrestricted
sine-Gordon model. Ground state \#1 is obtained by putting \protect\( \alpha =\frac{\pi }{3}\protect \),
while for ground state \#2 \protect\( \alpha =\frac{2\pi }{3}\protect \). The
TCS data shown have the predicted bulk energy term subtracted. }
\end{table}
To summarize, we now have sufficient evidence to believe that the \( \alpha  \)-twisted
NLIE describes the correct scaling functions for ground states of minimal models
perturbed by \( \Phi _{(1,3)} \) \emph{even in the nonunitary case}. However,
the NLIE for sine-Gordon is known to work for excited states as well. But how
do we get the excited state spectrum of the minimal models now?

\section{Excited states in minimal models perturbed by \protect\( \Phi _{(1,3)}\protect \)}

\subsection{The excited state equation}

In this section we write down the NLIE for excited states of Virasoro minimal
models perturbed by \( \Phi _{(1,3)} \) and we also point out some essential
differences from the results obtained by P. Zinn-Justin in \cite{zinn-justin}.

The \( \alpha  \)-twist can be recast in the language of the original light-cone
lattice Bethe Ansatz as a twist angle \( \omega  \) in the Bethe Ansatz equations
(BAE) which take the form \cite{omega-twist}
\begin{equation}
\label{bethe}
\left( \displaystyle\frac{\sinh \displaystyle\frac{\gamma }{\pi }\left[ \vartheta _{j}+\Theta +\displaystyle\frac{i\pi }{2}\right] \sinh \displaystyle\frac{\gamma }{\pi }\left[ \vartheta _{j}-\Theta +\displaystyle\frac{i\pi }{2}\right] }{\sinh \displaystyle\frac{\gamma }{\pi }\left[ \vartheta _{j}+\Theta -\displaystyle\frac{i\pi }{2}\right] \sinh \displaystyle\frac{\gamma }{\pi }\left[ \vartheta _{j}-\Theta -\displaystyle\frac{i\pi }{2}\right] }\right) ^{N}=-e^{-2i\omega }\prod _{k=1}^{M}\displaystyle\frac{\sinh \displaystyle\frac{\gamma }{\pi }\left[ \vartheta _{j}-\vartheta _{k}+i\pi \right] }{\sinh \displaystyle\frac{\gamma }{\pi }\left[ \vartheta _{j}-\vartheta _{k}-i\pi \right] }\: .
\end{equation}
 where 
\[
\gamma =\frac{\pi }{p+1}\]
and for the rest of the notations we refer the reader to \cite{DdV-97, our_npb}. 

From the BAE (\ref{bethe}) the NLIE for excited states can be derived by the
well-known methods \cite{DdV-97, our_npb}. We obtain the result 
\begin{equation}
\label{alpha-nlie}
\displaystyle \begin{array}{cc}
Z(\vartheta )=l\sinh \vartheta +g(\vartheta |\vartheta _{j})+\alpha  & -i\displaystyle\int ^{\infty }_{-\infty }dxG(\vartheta -x-i\eta )\log \left( 1+(-1)^{\delta }e^{iZ(x+i\eta )}\right) \\
 & +i\displaystyle\int ^{\infty }_{-\infty }dxG(\vartheta -x+i\eta )\log \left( 1+(-1)^{\delta }e^{-iZ(x-i\eta )}\right) 
\end{array}
\end{equation}
 The parameter \( \alpha  \) turns out to be
\begin{equation}
\label{alpha_giovanni}
\alpha =\omega \displaystyle\frac{p+1}{p}+\chi _{\infty }\left( \left[ \displaystyle\frac{1}{2}+\displaystyle\frac{S}{p+1}+\displaystyle\frac{\omega }{\pi }\right] -\left[ \displaystyle\frac{1}{2}+\displaystyle\frac{S}{p+1}-\displaystyle\frac{\omega }{\pi }\right] \right) 
\end{equation}
where
\[
\chi _{\infty }=\chi (\infty )=\frac{\pi }{2}\frac{p-1}{p}\]
and \( S \) is the Bethe spin as determined from eqn. (\ref{conteggi}). The
square brackets denote the integer part, i.e. \( [x] \) is the largest integer
smaller or equal to \( x \). The relation (\ref{alpha_giovanni}) can be obtained
in the following way. We recall that it is the derivative of the NLIE that can
be derived from the lattice Bethe Ansatz using a contour integral trick \cite{DdV-97, our_npb}.
Therefore it has to be integrated to give the NLIE itself and so there appears
an integration constant, which can be determined by matching the asymptotic
values of the lattice counting function \( Z \) coming from the NLIE to the
asymptotic values determined from the Bethe Ansatz \cite{DdV-97}. This is the
same procedure that yields the integration constant \( C \) in (\ref{nlie-cont}).
However, while \( C \) is a multiple of \( \pi  \) and can be absorbed by
redefining the counting function \( Z \) and the Bethe quantum numbers \( I_{j} \)
in (\ref{quantum}) (and possibly changing \( \delta  \) to \( 1-\delta  \)),
this is not true for \( \alpha  \) as it can take any real value in general.

In contrast, in \cite{zinn-justin} the form of \( \alpha  \) is 
\begin{equation}
\label{alpha_zinnjustin}
\alpha =\displaystyle\frac{\omega }{1-\displaystyle\frac{\gamma }{\pi }}=\omega \displaystyle\frac{p+1}{p}
\end{equation}
and the choice for \( \omega  \) is \( \gamma  \). However, this formula misses
the second term in eqn. (\ref{alpha_giovanni}). This term is extremely important
since it guarantees that the formula (\ref{alpha_giovanni}) satisfies the property
\[
\alpha \, \rightarrow \, \alpha +2\pi \, \, \, {\rm when}\, \, \, \omega \, \rightarrow \, \omega +\pi \, .\]
This relation is required for self-consistency: since the BAE (\ref{bethe})
are invariant under the shift of \( \omega  \) by \( \pi  \), the NLIE (which
is equivalent to the BAE before taking the continuum limit) must be invariant
under this shift as well. (Note that shifting \( \alpha  \) by \( 2\pi  \)
is an invariance of the NLIE (\ref{alpha-nlie}), with an appropriate redefinition
of \( Z \) and the Bethe quantum numbers \( I_{j} \) .)

From now on we restrict ourselves to the case of neutral (i.e. \( S=0 \)) states.
It is easy to see that even for states with a zero charge the relation between
\( \alpha  \) and \( \omega  \) is highly nontrivial.

Now we proceed to show that choosing 
\[
\omega =k\gamma \]
where \( k \) is integer, as the value of \( \omega  \) we can reproduce all
the required values of \( \alpha  \) listed in equation (\ref{twists}). First
of all, we substitute the value of \( p \) from (\ref{p_min_r_s}) to obtain
\[
\omega =\frac{k(s-r)\pi }{s}\, \, .\]
Since \( r \) and \( s \) are relative primes, the independent values of \( \omega \, \, \bmod \, \, \pi  \)
can be written as 
\[
\omega =\frac{l\pi }{s}\, \, ,\, \, l=0,\ldots ,s-1\, \, .\]
 For \( S=0 \), we can rewrite the formula (\ref{alpha_giovanni}) as follows
\[
\alpha =\frac{l\pi }{r}+\frac{2r-s}{2r}\pi \left( \left[ \frac{1}{2}+\frac{l}{s}\right] -\left[ \frac{1}{2}-\frac{l}{s}\right] \right) \, .\]
We are interested only in the value of \( \alpha \, \, \bmod \, \, \pi  \),
since using the parameter \( \delta  \) one can effectively shift \( \alpha  \)
by \( \pi  \). This leaves us with the formula 
\[
\alpha =\frac{l\pi }{r}-\frac{s\pi }{2r}\left( \left[ \frac{1}{2}+\frac{l}{s}\right] -\left[ \frac{1}{2}-\frac{l}{s}\right] \right) \, .\]
The first possibility is that \( l<\frac{s}{2} \), which simply gives us the
values 
\[
\alpha =\frac{l\pi }{r}.\]
 When \( s \) is even, we can have \( l=\frac{s}{2} \), which gives us \( \alpha =0 \).
Finally, when \( l>\frac{s}{2} \), we get the values 
\[
\alpha =\frac{(l-s)\pi }{r}\, \, .\]
It is easy to check that these formulae reproduce every value 
\[
\alpha =\frac{n\pi }{r}\, \, \bmod \, \, \pi \]
at least once, using the fact that \( s>r \) and that the values above form
an uninterrupted sequence of \( s \) numbers (or when \( s \) is even, of
\( s-1 \) numbers, the zero repeated) with equal distances \( \frac{\pi }{r} \).

As we have already seen in the previous section, all the values 
\[
\alpha =\frac{k\pi }{r}\, \, ,\, \, k=1,\ldots ,r-1\]
are necessary to reproduce correctly the \( r-1 \) ground states of the model
\( Vir(r,s)+\Phi _{(1,3)} \). The single choice 
\[
\alpha =\frac{\pi }{p}\]
of the paper \cite{zinn-justin} is not enough to reproduce all the possible
states. The twisted lattice Bethe Ansatz was analyzed by de Vega and Giacomini
in \cite{omega-twist}. On the lattice, passing from the sine-Gordon model to
the perturbed Virasoro model amounts to going from the six-vertex model to a
lattice RSOS model. In \cite{omega-twist} it was shown that to obtain all the
states of the RSOS model it is necessary to take all the twists 
\[
\omega =k\gamma \, \, \bmod \, \, \pi \]
into account. The fact that not all these twists correspond to inequivalent
values of \( \alpha  \) and so to different physical states is a consequence
of the RSOS truncation.

To close this section we remark that the parameter \( \alpha  \) drops out
of the second determination of \( Z \) (\ref{nlie-2nd}) in the attractive
regime. This is important because as a consequence the IR asymptotics of the
breather states does not depend on \( \alpha  \) and so the \( S \)-matrices
involving breathers are unchanged. In fact, scattering amplitudes between solitons
and breathers remain unchanged too, as can be seen from examining the argument
that we used to derive them in section 2. This matches with the fact that the
RSOS restriction from sine-Gordon theory to perturbed minimal models does not
modify scattering amplitudes that involve two breathers or a breather and soliton
\cite{reshetikhin-smirnov}.

\subsection{The UV limit}

There is in fact a very simple intuitive argument to show that the states we
get from (\ref{alpha-nlie}) are related to the minimal models in the UV limit.
To keep the formulae simple, we present it here for the \emph{unitary} case
\( p=r=s-1 \). Let us first recall that the UV limit of the original NLIE (\ref{nlie-cont})
yields the vertex operators \( V_{(n,m)} \) and their descendants, where for
general choice of \( \delta  \), \( n \) is a half-integer and \( m \) is
an integer. 

Let us look at the weights in the neutral sector, which means \( m=0 \). Using
the formulas for the UV limit of the NLIE from our previous paper \cite{our_npb},
one can see that introducing \( \alpha \neq 0 \) is equivalent to shifting
the quantum number \( n \) to \( n+\frac{\alpha }{2\pi } \). One has to be
careful that since the value of the central charge is shifted from \( 1 \)
to the one of the minimal model, we have to take this shift into account when
computing the conformal weight from the leading UV behaviour of the energy level
(\ref{energy_UV}). We put in the value 
\[
\alpha =\frac{l\pi }{p},\, \, k=1,\ldots ,r-1\, \, ,\]
and the resulting conformal weights take the form
\begin{equation}
\label{alpha_weights}
\Delta ^{+}=\Delta ^{-}=\displaystyle\frac{(2np+l)^{2}-1}{4p(p+1)}\, \, .
\end{equation}
Comparing to the formula (\ref{Kac_formula}) we see that this is the weight
of the field \( \Phi _{(l,l-2n)} \) in the minimal model \( Vir(p,p+1) \).
Therefore one expects that we get something related to minimal models, however,
in order not to overflow the Kac table, the range of \( n \) must be restricted
as 
\[
1\leq l-2n\leq p\, \, .\]
For charged states, a similar calculation can be performed. We find in particular
that
\begin{equation}
\label{fractional_spin}
2\left( \Delta ^{+}-\Delta ^{-}\right) =m\displaystyle\frac{\alpha }{\pi }\, \, \bmod \, \, 1\, ,
\end{equation}
so these states have fractional Lorentz spin in general.

In a more precise way, the UV limit of the twisted NLIE can be examined with
exactly the same method as outlined in \cite{DdV-97, our_npb}, so here we just
give the formulae necessary to do the computations and for the derivation and
notational details we refer to the papers above. The \emph{right/left kink equation}
can be obtained by substituting 
\[
\vartheta \, \, \rightarrow \, \, \vartheta \pm \log \frac{2}{l}\]
 into the NLIE and keeping only the leading terms when \( l\, \, \rightarrow \, \, 0 \).
It takes the form 
\begin{equation}
\label{kink}
\displaystyle Z_{\pm }(\vartheta )=\pm e^{\pm \vartheta }+\alpha +g_{\pm }(\vartheta )+\displaystyle\int ^{\infty }_{-\infty }dxG(\vartheta -x){\cal Q}_{\pm }(x)\: ,
\end{equation}
where the source term \( g_{\pm }(\vartheta ) \) is the limit 
\begin{equation}
\label{kink_source}
g_{\pm }(\vartheta )=g\left( \vartheta \pm \log \displaystyle\frac{2}{l}\right) \, \, ,\, \, l\, \rightarrow \, 0\, \, ,
\end{equation}
and 
\[
{\cal Q}_{\pm }(x)=\lim _{\eta \rightarrow +0}-i\log \frac{1+(-1)^{\delta }e^{iZ(x+i\eta )}}{1+(-1)^{\delta }e^{-iZ(x-i\eta )}}\]
We call a source \emph{right/left moving} if its position given in this limit
by
\[
\vartheta ^{\pm }_{j}\pm \log \frac{2}{l}\, \, ,\]
where \( \vartheta _{j}^{\pm } \) is finite as \( l\, \, \rightarrow \, \, \infty  \).
We denote the number of right/left-moving holes by \( N^{\pm }_{H} \) and their
positions by \( h^{\pm }_{j} \) and similarly introduce the notations \( N^{\pm }_{S}\, \, ,\, \, y^{\pm }_{j} \)
for the special objects, \( M^{\pm }_{C}\, \, ,\, \, c^{\pm }_{j} \) for the
close complex roots, and \( M^{\pm }_{W}\, \, ,\, \, w^{\pm }_{j} \)for the
wide complex roots. There can also be sources whose positions remain finite
as \( l\, \, \rightarrow \, \, \infty  \) : they are called \emph{central}.
We introduce \( S^{\pm } \) as 
\[
S^{\pm }=\frac{1}{2}[N_{H}^{\pm }-2N_{S}^{\pm }-M_{C}^{\pm }-2M^{\pm }_{W}\theta (p-1)]\, \, .\]
 The \emph{plateau equation} takes the form:
\begin{equation}
\label{plateau}
Z_{\pm }(\mp \infty )=\alpha +g_{\pm }(\mp \infty )+\displaystyle\frac{\chi _{\infty }}{\pi }\omega _{\pm }\, \, ,\, \, \, \chi _{\infty }=\displaystyle\frac{\pi }{2}\displaystyle\frac{p-1}{p}\, \, ,
\end{equation}
where we have
\[
Z_{\pm }(\mp \infty )=\omega _{\pm }+\pi \delta +2\pi k_{\pm }\: \, ,\, \, \, -\pi \leq \omega _{\pm }\leq \pi \, \, ,\]
and 
\[
g_{\pm }(\mp \infty )=\pm 2\left( S-2S^{\pm }\right) \chi _{\infty }+2\pi k^{\pm }_{W}\: .\]
The numbers \( k^{\pm }_{W} \) depend on the configuration of wide roots and
are generally integers except when we have an odd number of self-conjugate roots,
in which case they are half-integers. From these relations one can compute the
value of \( \omega _{\pm } \) to obtain
\[
\omega _{\pm }=\frac{2\alpha p}{p+1}\pm 2\pi \left( S-2S^{\pm }\right) \frac{p-1}{p+1}+4\pi \frac{p}{1+p}k_{\pm }\, \, ,\]
where \( k_{\pm } \) are some (half)-integers. The conformal weights themselves
can be computed using the formula 
\begin{equation}
\label{delta}
\Delta _{\pm }=\displaystyle\frac{c-1}{24}\pm \left( I^{\pm }_{H}-2I^{\pm }_{S}-I^{\pm }_{C}-I^{\pm }_{W}\right) +\displaystyle\frac{\Sigma _{\pm }}{2\pi }+\displaystyle\frac{p+1}{p}\displaystyle\frac{\omega _{\pm }^{2}}{16\pi ^{2}}\: ,
\end{equation}
where \( I^{\pm }_{H} \) is the sum of the Bethe quantum numbers of left/right
holes etc., \( c \) is the central charge
\[
c=1-\frac{6(r-s)^{2}}{rs}\, \, ,\]
and the quantities \( \Sigma _{\pm } \) are given by
\[
\Sigma _{\pm }=-\sum ^{N^{\pm }_{H}}_{j=1}g_{\pm }\left( h_{j}^{\pm }\right) +2\sum ^{N^{\pm }_{S}}_{j=1}g_{\pm }\left( y_{j}^{\pm }\right) +\sum ^{M^{\pm }_{C}}_{j=1}g_{\pm }\left( c_{j}^{\pm }\right) +\sum ^{M^{\pm }_{W}}_{j=1}g_{\pm }\left( w_{j}^{\pm }\right) _{II}\mp 2S^{\pm }\alpha \, \, ,\]
which can be written more explicitly as follows
\begin{equation}
\label{sigmapm}
\Sigma _{\pm }=\mp 2S^{\pm }\alpha -4S^{\pm }(S-S^{\pm })\chi _{\infty }+2\pi q^{\pm }_{W}\, ,
\end{equation}
where \( q_{W}^{\pm } \) is an integer or half-integer which depends on the
configuration of wide roots. One has to be careful that \( g_{\pm }(\vartheta )_{II} \)
is defined by 
\begin{equation}
\label{kink_source_2nd}
g_{\pm }(\vartheta )_{II}=g\left( \vartheta \pm \log \displaystyle\frac{2}{l}\right) _{II}\, \, ,\, \, l\, \rightarrow \, 0\, \, ,
\end{equation}
and \emph{not} as the second determination of \( g_{\pm }(\vartheta ) \).

We close this section with some immediate consequences of the above formulae.
First note that because the value of \( \alpha  \) for a minimal model is never
a multiple of \( \pi  \), one does not expect central sources in the UV limit
(in all the examples of \cite{our_npb} with central sources, the left-right
symmetry of the NLIE was crucial. This symmetry, however, only holds for \( \alpha =0 \)
or \( \pi  \)). As a consequence we have 
\[
S=S^{+}+S^{-}\, \, ,\]
and in addition \( \omega _{+}=\omega _{-} \), which means that there are only
\emph{one-plateau systems}. By an inspection of the formula (\ref{delta}) this
implies that for any state with \( S=0 \)
\[
\Delta ^{+}-\Delta ^{-}\]
is integer or half-integer. In fact, choosing the quantization rule and the
parameter \( \delta  \) in an appropriate way, one can ensure that this difference
is integer (see \cite{our_npb, our_letter2}). This means that the UV limit
of any neutral state is either a field occurring in the ADE classification of
modular invariants \cite{ADE} or (in case we choose \( \delta  \) so that
\( \Delta ^{+}-\Delta ^{-} \) is half-integer) it is a field from a fermionic
version of the minimal model \cite{heretic}. 

For the case of neutral states, an elementary calculation gives the following
formula
\[
\Delta ^{+}=\Delta ^{-}=\frac{c-1}{24}+\frac{\left( \frac{\alpha }{\pi p}+\left( 2k-4S^{+}\right) p\right) ^{2}}{4p(p+1)}+N_{\pm }\, \, ,\]
where \( k \), \( N_{\pm } \) are integer or half-integer which upon substituting
the value 
\[
\alpha =\frac{l\pi }{p}\]
is just identical to the formula (\ref{alpha_weights}), apart from \( N_{\pm } \),
which must in fact be nothing else but descendant numbers and so one expects
that they are eventually integer. A similar result was obtained in \cite{zinn-justin},
but just as we said above the choice made in that paper does not allow one to
fill the Kac table: for unitary models it produces only the fields \( \Phi _{(1,n)} \).
However, similarly to the argument presented at the beginning of this section,
it is not clear whether the weights actually stay inside the Kac table, for
which in the unitary case one must require
\[
1<l-2k+4S^{+}<p\, .\]
Due to the fact that the configuration of sources in the UV may be very non-trivially
related to the one in the IR, this condition is very hard to check in general,
but no concrete examples that we calculated have ever violated this bound.

From the formulae (\ref{delta}) and (\ref{sigmapm}) it is also clear that
in general 
\[
2\left( \Delta ^{+}-\Delta ^{-}\right) =2S\frac{\alpha }{\pi }\, \, \bmod \, \, 1,\]
and so we see again that general charged states will have fractional Lorentz
spin. Remembering the relation \( m=2S \) \cite{our_npb}, this agrees with
our previous result (\ref{fractional_spin}). Actually, it is known that charged
states in the models \( Vir(r,s)+\Phi _{(1,3)} \) generally have fractional
Lorentz spin \cite{felder-leclair}.

\section{Concrete examples of excited states}

\subsection{The \protect\( Vir(2,2n+1)+\Phi (1,3)\protect \) series}

Let us start with examining the scaling Lee-Yang model \( Vir(2,5)+\Phi _{(1,3)} \).
There is only one independent value of the twist which we choose as 
\[
\alpha =\frac{\pi }{2}\, \, ,\]
since we have a single ground state, and as a result there are no kinks in the
spectrum. We fix the value of \( \alpha  \) as above, so we still have a freedom
of choosing \( \delta  \). This can be done by matching to the UV dimensions:
if for a certain state we choose the wrong value of \( \delta  \), we find
a conformal dimension that is not present in the Kac table of the model. 

The excited states are multi-particle states of the first breather of the corresponding
unrestricted sine-Gordon model, which has 
\[
p=\frac{2}{3}.\]
Now one can calculate the state containing one particle at rest. We find the
numerical data presented in table \ref{yl_first}. 

It turns out that as we decrease \( l \), the self-conjugate root starts moving
to the right. It does not remain in the middle like in the \( \alpha =0 \)
case, which is to be expected since for nonzero \( \alpha  \) we have no left/right
symmetry. However, the total momentum of the state still remains zero due to
a contribution from the integral term in eqn. (\ref{momentum}).
\begin{table}
{\centering \begin{tabular}{|c|c|c|}
\hline 
\( l_{B} \)&
TCS&
NLIE\\
\hline 
0.1&
23.05277&
n/a\\
\hline 
0.5&
4.679779&
n/a\\
\hline 
1.0&
2.447376&
n/a\\
\hline 
1.5&
1.748874&
n/a\\
\hline 
2.0&
1.430883&
n/a\\
\hline 
2.6&
1.238051&
1.238012(\#)\\
\hline 
3.0&
1.164321&
1.164319\\
\hline 
3.5&
1.105220&
1.105196\\
\hline 
4.0&
1.068256&
1.068237\\
\hline 
5.0&
1.029356&
1.029348\\
\hline 
\end{tabular}\par}

\caption{\small \label{yl_first} The first excited state of the scaling Lee-Yang model. The
energy and the volume are normalized to the mass of the lowest excitation, which
is the first breather of the unrestricted sine-Gordon model. The TCS data shown
have the predicted bulk energy term subtracted. }
\end{table}
One can see that once again we have the phenomenon noticed in the case of the
first breather of sine-Gordon theory, namely the appearance of the special root
and its two accompanying holes, so the iteration breaks down again around \( l=2.5 \).
The (\#) in the table \ref{yl_first} written after the NLIE result for \( l=2.6 \)
means that due to the fact that the singularities corresponding to the new holes
and the special root are just about to cross the contour and upset the iteration
scheme, the NLIE result becomes less precise. We will use this notation on later
occasions too. In any case, the agreement still looks quite convincing. 

Let us now look at the UV spectrum of the model. We know that the Lee-Yang model
contains only two primary fields, the identity \( \mathbb I \) and the field
\( \varphi  \) with left/right conformal weights
\[
\Delta ^{+}=\Delta ^{-}=-\frac{1}{5}\, \, .\]
In fact, the ground state of the massive model corresponds to \( \varphi  \)
in the UV limit. One can compute the UV limit of the first particle from the
NLIE too, taking into account the appearance of the special root and the holes.
It turns out that the special root and one of the holes moves to the left together
with the self-conjugate root, while the other hole moves to the right. The result
is
\[
\Delta ^{+}=\Delta ^{-}=0\, \, ,\]
i.e. the identity operator \( \mathbb I \), which fits nicely with the TCS
data (see also \cite{yurov-zamolodchikov}).

Let us look now at moving breathers. If the self-conjugate root has Bethe quantum
number \( I=1 \), the corresponding state will have momentum quantum number
\( 1 \), i.e. 
\[
P=\frac{2\pi }{R}\, \, ,\]
and in the UV \( \Delta ^{+}-\Delta ^{-}=1 \). One can note from the numerical
data presented in table \ref{yl_first_spin1} that the special root does not
appear here. The reason is that the self-conjugate root moves to the left and
the real part of its position \( \vartheta  \) is given to leading order by
\[
\sinh (\Re e\, \vartheta )\, \sim \, -\frac{2\pi I}{l_{B}}\, \, .\]
As a result, the contribution to the derivative of \( Z \) from the \( l\sinh \vartheta  \)
term remains finite when \( l\, \rightarrow \, 0 \). In the previous example
of the particle at rest the left-moving nature of the self-conjugate root when
\( I=0 \) does not prevent the occurrence of the breakdown in the iteration
scheme: since its Bethe quantum number is zero, it does not move fast enough
to the left in order to balance the negative contribution to derivative of \( Z \)
coming from the self-conjugate root source. At the moment we have no way of
predicting analytically whether or not there will be specials in the UV limit:
we just use the numerical results to establish the configuration for the evaluation
of UV weights, supplemented with a study of the self-consistency of the solution
of the plateau equation (\ref{plateau}).
\begin{table}
{\centering \begin{tabular}{|c|c|c|}
\hline 
\( l_{B} \)&
TCS&
NLIE\\
\hline 
0.1&
60.75048&
60.74682\\
\hline 
0.5&
12.20618&
12.20561\\
\hline 
1.0&
6.182516&
6.182363\\
\hline 
1.5&
4.202938&
4.202915\\
\hline 
2.0&
3.231734&
3.231640\\
\hline 
2.5&
2.662186&
2.662110\\
\hline 
3.0&
2.292273&
2.292231\\
\hline 
3.5&
2.035552&
2.035530\\
\hline 
4.0&
1.848892&
1.848849\\
\hline 
5.0&
1.599792&
1.599762\\
\hline 
\end{tabular}\par}

\caption{\small \label{yl_first_spin1} The one-particle states with Lorentz spin 1 of the
scaling Lee-Yang model. The energy and the volume are normalized to the mass
of the lowest excitation, which is the first breather of the unrestricted sine-Gordon
model. The TCS data shown have the predicted bulk energy term subtracted. }
\end{table}
The UV dimensions for the moving breather turn out to correspond to the state
\( L_{-1}\varphi  \).

One can similarly compute the UV dimensions for some other excited states. For
example, the two-particle states with half-integer Bethe quantum numbers \( I_{1}>0,\, \, I_{2}<0 \)
for the two self-conjugate roots are found to have 
\[
\Delta ^{+}=-\frac{1}{5}+I_{1}+\frac{1}{2}\, \, ,\, \, \Delta ^{-}=-\frac{1}{5}-I_{2}+\frac{1}{2}\, \, ,\]
in agreement with TCS data which show that they correspond in the UV to descendent
states of \( \varphi  \). The first such state with quantum numbers 
\[
I_{1}=\frac{1}{2}\, \, ,\, \, I_{2}=-\frac{1}{2}\]
corresponds in the UV to \( L_{-1}\bar{L}_{-1}\varphi  \) and is given numerically
in table \ref{yl_second}.

\begin{table}
{\centering \begin{tabular}{|c|c|c|}
\hline 
\( l \)&
TCS&
NLIE\\
\hline 
0.1&
123.583&
123.5693\\
\hline 
0.5&
24.7806&
24.77936\\
\hline 
1.0&
12.4870&
12.48635\\
\hline 
1.5&
8.42926&
8.428693\\
\hline 
2.0&
6.42931&
6.429201\\
\hline 
3.0&
4.48444&
4.484209\\
\hline 
4.0&
3.56367&
3.563519\\
\hline 
5.0&
3.04899&
3.048881\\
\hline 
\end{tabular}\par}

\caption{\small \label{yl_second} The lowest lying zero-momentum two-particle state in the
scaling Lee-Yang model as computed from the NLIE and compared with TCS.}
\end{table}
The lowest lying three-particle state of zero momentum, with Bethe quantum numbers
\( (-1,0,1) \) corresponds to the left/right symmetric second descendent of
the identity field, i.e. to the field \( T\bar{T} \), where \( T \) denotes
the energy-momentum tensor. This is very interesting, since from experience
with NLIE UV calculations one would naively expect this to be a first descendent
(descendent numbers are usually linked to the sum of Bethe quantum numbers of
left/right moving particles - see the examples in \cite{our_npb} -, and this
state is the lowest possible descendent of the identity \( \mathbb I \)). However,
the field \( L_{-1}\bar{L}_{-1}\mathbb I \) is well-known to be a null field
in any conformal field theory.

The above correspondences are again confirmed by comparing to TCS (see the wonderful
figures in \cite{yurov-zamolodchikov}). In general, one can establish the rule
that states with odd number of particles must be quantized by integers (\( \delta =1 \)),
while those containing even number of particles must be quantized by half-integers
(\( \delta =0 \)) in order to reproduce correctly the spectrum of the scaling
Lee-Yang model.

We conducted similar studies for the models \( Vir(2,7)+\Phi _{(1,3)} \) and
\( Vir(2,9)+\Phi _{(1,3)} \) and found similarly good agreement with TCS data.
For the first one-particle state of the model \( Vir(2,7)+\Phi _{(1,3)} \)
we also checked our results against the TBA data in the numerical tables of
\cite{dorey_tateo2} and found agreement with the TBA results.

Given the choice of \( \alpha  \) above, the correct rule of quantization in
all of the models \( Vir(2,2n+1)+\Phi _{(1,3)} \) is 
\[
\delta =M_{sc}\, \, \bmod \, \, 2\, \, ,\]
where \( M_{sc} \) is the number of self-conjugate roots in the source corresponding
to the state. This is exactly the same rule as the one established for pure
sine-Gordon theory in \cite{our_npb}. In the presence of the twist, such a
rule of course has meaning only together with a definite convention for the
choice of \( \alpha  \).

\subsection{One-breather states in the \protect\( Vir(3,7)\protect \) case}

It is interesting to note that in the case of \( Vir(3,n) \) models, all the
neutral states must come in two copies, since they can be built on top of either
of the two ground states. We take the example of the \( Vir(3,7) \) model and
the states corresponding to a breather at rest. We have 
\[
p=\frac{3}{4}\]
 but now there are two inequivalent values for the twist 
\[
\alpha =\frac{\pi }{3}\, \, ,\, \, \frac{2\pi }{3}\, .\]
When \( \alpha =0 \), we can calculate the critical value of \( l \) to be
\( l_{critical}=5.23 \) using (\ref{critical_estimate}). In this case, the
twist helps a bit, because it makes the self-conjugate root a left mover; it
is intuitively clear that the bigger the twist, the more it lowers the eventual
value of \( l_{crtical} \), which is in accord with the numerical results of
table \ref{37_firsts}. From the TCS data one can identify that breather \#1
is really the one-particle state in the sector of ground state \#1, while breather
\#2 is in the sector over ground state \#2 using the notations of table \ref{vacua_37}.

A direct calculation of the conformal weights gives the following results:
\[
\Delta ^{+}=\Delta ^{-}=\frac{3}{28}\]
for breather \#1 and 
\[
\Delta ^{+}=\Delta ^{-}=0\]
for breather \#2, which are in complete agreement with the TCS data. We also
checked the two different states containing two breathers with Bethe quantum
numbers
\[
I_{1}=\frac{1}{2}\, \, ,\, \, I_{2}=-\frac{1}{2}\, \, ,\]
and found an equally excellent numerical agreement with TCS. Just like in the
case of sine-Gordon and scaling Lee-Yang model, for these states one can continue
the iteration of the NLIE down to any small value of \( l \), although at the
expense of a growing number of necessary iterations to achieve the prescribed
precision.
\begin{table}
{\centering \begin{tabular}{|c|c|c|c|c|}
\hline 
&
\multicolumn{2}{|c|}{ breather 1}&
\multicolumn{2}{|c|}{breather 2}\\
\hline 
\( l \)&
TCS&
NLIE&
TCS&
NLIE\\
\hline 
0.1&
32.21645&
n/a&
18.76030&
n/a\\
\hline 
0.5&
6.671964&
n/a&
4.037716&
n/a\\
\hline 
1.0&
3.662027&
n/a&
2.434501&
2.434431(\#)\\
\hline 
1.5&
2.769403&
n/a&
2.027227&
2.027213\\
\hline 
2.0&
2.385451&
n/a&
1.884404&
1.884388\\
\hline 
2.5&
2.190232&
n/a&
1.829459&
1.829456\\
\hline 
3.0&
2.079959&
n/a&
1.809244&
1.809248\\
\hline 
3.5&
2.012596&
n/a&
1.803873&
1.803886\\
\hline 
4.0&
1.968808&
1.968784(\#)&
1.804937&
1.804953\\
\hline 
4.5&
1.938895&
1.938889&
1.808633&
1.808658\\
\hline 
5.0&
1.917648&
1.917635&
1.813191&
1.813224\\
\hline 
\end{tabular}\par}

\caption{\small \label{37_firsts} The two one-breather states of the Virasoro minimal model
\protect\( Vir(3,7)\protect \) perturbed by \protect\( \Phi _{(1,3)}\protect \).
The energy and the volume are normalized to the mass of the kink, which is the
soliton of the unrestricted sine-Gordon model. Breather \#1 has \protect\( \alpha =\frac{\pi }{3}\protect \),
while breather \#2 corresponds to \protect\( \alpha =\frac{2\pi }{3}\protect \).
The TCS data shown have the predicted bulk energy term subtracted. }
\end{table}

\section{Conclusions}

In this paper we described how to use the NLIE approach to compute finite size
effects in Virasoro minimal models perturbed by \( \Phi _{(1,3)} \). Up to
now, apart from a previous attempt by P. Zinn-Justin \cite{zinn-justin}, the
only results in this framework were the description of ground states of \( \Phi _{(1,3)} \)
perturbed unitary minimal models \cite{fioravanti}.

As a starting point, we first extended our previous studies \cite{our_letter1, our_npb, our_letter2}
of sine-Gordon theory to breather states which then played a prominent role
in providing us concrete examples of excited states in minimal models. We showed
that in the IR limit the NLIE successfully reproduces the scattering amplitudes
involving breathers and we gave examples of comparing numerical results from
the NLIE to those coming from TCS. 

We then proceeded to the case of perturbed minimal models. Using a twisted NLIE
{\'a} la Zamolodchikov \cite{plymer} we provided examples for ground states
in nonunitary models, thereby overcoming the limitation of \cite{fioravanti}
which considered only the unitary case. 

Concerning excited states, we started by establishing the connection of the
twisted NLIE to a twisted version of the lattice Bethe Ansatz. Doing so, we
have found two shortcomings of the results in \cite{zinn-justin}. The first
of these was an incorrect relation (\ref{alpha_zinnjustin}) between the twist
parameter \( \omega  \) in the lattice Bethe Ansatz and the one appearing in
the NLIE, which did not reflect the periodicity of the Bethe Ansatz in the \( \omega  \),
in contrast to the relation (\ref{alpha_giovanni}) found by us. The other was
that to describe all possible states of the UV limiting minimal CFT, it is not
enough to choose just one value of the twist parameter: indeed a whole range
of values is required (\ref{twists}). We demonstrated that the twisted NLIE
gives conformal weights that are consistent with the spectrum of minimal models.
Numerical calculation of concrete examples gave us a strong evidence for the
correctness of the energy levels derived from the twisted NLIE for excited states.

Throughout this paper we made one important omission: we did not treat the multi-kink
states so characteristic of the perturbed minimal models. Although the general
treatment of the UV spectrum in section 4.2 is valid for those states too, a
detailed description is far more complicated than for states which contain only
breathers and is left open to further studies.

We also pointed out a technical difficulty, namely that the source configuration
of the NLIE may change as we vary the volume parameter \( l \). Typically what
happens is that while the counting function \( Z \) is monotonic on the real
axis for large volume, this may change as we lower the value of \( l \) and
so-called special sources (and accompanying holes) may appear. We do not as
yet have any consistent and tractable numerical iteration scheme to handle this
situation, although the analytic UV calculations and intuitive arguments show
that the appearance of these terms in the NLIE is consistent with all expectations
coming from the known properties of perturbed CFT. In addition, in the range
of \( l \) where we can iterate the NLIE without difficulty, our numerical
results show perfect agreement with TCS. We want to emphasize that these transitions
are not physical: the counting function \( Z \) and the energy of the state
is expected to vary analytically with the volume, it is just described by a
NLIE with modified source terms. As it was pointed out in \cite{DdV-97, our_npb},
the whole issue is related to the choice of the branch of the logarithmic term
in the NLIE (\ref{nlie-cont}). The problem itself is very similar to the behaviour
of singularities encountered in the study of the analytic continuation of the
TBA equation \cite{tateo_dorey} and we can hope that establishing a closer
link between the two approaches can help to clarify the situation.

We think that the work presented in this paper provides a strong evidence for
the NLIE description of excited states in perturbed minimal models. We would
like to mention that following the lines of \cite{zinn-justin} it is possible
to extend this framework to minimal models of \( W \) algebras based on the
\( ADE \) Lie algebras. It seems very likely that apart from the loopholes
(concerning the twist parameters) pointed out and clarified in this paper no
further complications will occur, but a detailed discussion of these issues
is out of the scope of the present publication.

\bigskip{}
{\par\centering \textbf{Acknowledments}\par}
\medskip{}

We are indebted to P.E. Dorey for useful discussions. This work was partially
supported by European Union TMR Network FMRX-CT96-0012 and by INFN \emph{Iniziativa
Specifica} TO12. G. T. has been supported by an INFN postdoctoral fellowship
and partially by the FKFP 0125/1997 and OTKA T016251 Hungarian funds.

\end{document}